\documentclass[onecolumn,floatfix,superscriptaddress,a4paper,
               showpacs,showkeys,nofootinbib,preprint]{revtex4}
\usepackage{epsfig}
\usepackage{latexsym}
\usepackage{xspace}
\usepackage{hyperref}
\usepackage[latin2]{inputenc}
\usepackage{indentfirst}
\usepackage{enumerate}

\usepackage{amsmath}
\usepackage{amssymb}
\usepackage[english]{babel}
\usepackage{url}

\newcommand{\eq}[1]{\begin{align} #1 \end{align}}

\begin{document}

\title{Strongly Intensive Measures for Particle Number Fluctuations:
\\  Effects of Hadronic Resonances
       }

\author{Viktor V. Begun }
\affiliation{Institute of Physics, Jan Kochanowski University, Kielce, Poland}
\affiliation{Bogolyubov Institute for Theoretical Physics, Kiev, Ukraine}
\author{Mark I. Gorenstein}
\affiliation{Bogolyubov Institute for Theoretical Physics, Kiev, Ukraine}
 \affiliation{Frankfurt Institute for Advanced Studies, Frankfurt, Germany}
\author{Katarzyna Grebieszkow}
\affiliation{Faculty of Physics, Warsaw University of Technology, Warsaw, Poland}

\date{\today}

\begin{abstract}
Strongly intensive measures $\Delta$ and
$\Sigma$ are used to study event-by-event fluctuations
of hadron multiplicities in nucleus-nucleus collisions.
The effects of resonance decays are investigated within statistical model
and relativistic transport model.
Two specific examples are considered:
resonance decays to two positively charged
particles
(e.g., $\Delta^{++}\rightarrow p+ \pi^+$)
and to $\pi^+\pi^-$-pairs.
(e.g., $\rho^0\rightarrow \pi^-+\pi^+$).
It is shown that resonance abundances at the chemical freeze-out
can be estimated by measuring the fluctuations of the number of stable hadrons.
These model results are compared to the full hadron-resonance
gas analysis within both the grand canonical and canonical ensemble.
The ultra-relativistic quantum molecular dynamics (UrQMD) model of nucleus-nucleus
collisions is
used to illustrate the role of global charge conservation,
centrality selection, and limited experimental
acceptance.

\end{abstract}

\pacs{12.40.-y, 12.40.Ee}

\keywords{}

\maketitle
\section{Introduction}
The main physics motivation for the experimental investigations of relativistic
nucleus-nucleus (A+A) collisions started at mid 1980s was to create and study the
strongly interacting matter in its different phases: the hadron-resonance gas
and the quark-gluon plasma.  Today these studies
are still in progress at the Super Proton
Synchrotron (SPS) of the European Organization for Nuclear
Research (CERN) and at the Relativistic
Heavy Ion Collider (RHIC) of
Brookhaven National Laboratory (BNL); they
are pursued also at much higher collision energies at the
Large Hadron Collider (LHC) at CERN.
A possibility to observe signatures
of the critical point of the QCD matter inspired the energy and system size scan
program of the NA61/SHINE Collaboration at the CERN SPS \cite{Ga:2009}
and the low energy scan program
of the STAR and PHENIX Collaborations at the BNL RHIC
\cite{RHIC-SCAN}.

Experimental and theoretical investigations of the event-by-event
(e-by-e) fluctuations in A+A collisions
are relevant for  current and future studies of
the onset of deconfinement and the search for the critical point
(see, e.g., recent review \cite{GGS} and references therein).
These investigations, however, have been confronted with a serious problem.
The e-by-e fluctuations of  the number
of nucleon participants  affect strongly the
fluctuations of any physical observables
\cite{KGBG:2010}. In the language of statistical mechanics,
this is equivalent to the system volume fluctuations.
Note that in high energy A+A collisions the system volume fluctuations
can be hardly avoided. Besides, the average volume  of
the created matter and its variations from collision to collision are usually
difficult or even impossible to be measured. Therefore, a choice of adequate
statistical tools is crucially important for a study of the e-by-e fluctuations
in A+A collisions.

In the present paper we use the {\it strongly intensive} measures
of fluctuations  \cite{GG:2011}.
In a framework of several popular models of A+A collisions,
these quantities are independent of both the average volume
of a system and the volume fluctuations. For example, this is valid
within the grand canonical formulation of statistical mechanics.
An analysis of the yields of different hadronic species in A+A
collisions demonstrates that a statistical model of the hadron-resonance gas
gives an impressive agreement with a large amount of data
in terms of a few adjusting parameters.
Using statistical approach the effects of resonance decays
for the particle number fluctuations and correlations are considered
in terms of the strongly intensive measures.
We discuss two examples for which the e-by-e fluctuations of hadron multiplicities
are rather sensitive to the abundances of
resonances at the chemical freeze-out.
As a result, these resonance abundances, which are difficult to be measured by other methods,
can be estimated by measuring the fluctuations and correlations of the numbers of stable hadrons.
Note that an idea to use the e-by-e fluctuations
of particle number ratios to estimate the number of hadronic resonances was
suggested for the first time by Jeon and Koch in Ref.~\cite{JK}.

The paper is organized as follows. In Sec.~\ref{SI}
the notions of strongly intensive measures  and
volume fluctuations in thermal systems are considered. Sec.~\ref{sec-AM}
presents two simple examples of analytical calculations:
first, resonance decays
into two positive hadrons and, second,  into $\pi^+\pi^-$-pairs.
In Sec.~\ref{sec-HG}  these two model examples are tested within the
full hadron-resonance gas model.
In Sec.~\ref{sec-UrQMD}
the ultra-relativistic quantum molecular dynamics (UrQMD)
model is used in Pb+Pb and proton-proton ($p+p$) collisions at the SPS energies
to illustrate the role of  global charge conservation, centrality selection, and
limited experimental acceptance.
A summary in Sec.~\ref{sum} closes the article.

\section{Strongly Intensive Quantities}\label{SI}
The strongly intensive quantities $\Delta$ and $\Sigma$ have been introduced in
Ref.~\cite{GG:2011}. Within the grand canonical ensemble (GCE)
formulation of statistical mechanics they are independent of the
average volume and volume fluctuations. Similar properties take
place in the model of independent sources: the strongly intensive
measures of fluctuations are independent of the average number of
sources and of fluctuations of the number of sources.

Note that the first strongly intensive measure of fluctuations, the
so-called  $\Phi$ measure, was introduced a long time ago in
Ref.~\cite{GM:1992}. There were many attempts  to use the $\Phi$
measure in the data analysis~\cite{Phi_data,d1,d2,d3,d4,d5,d6} and
in theoretical
models~\cite{Phi_models,m2,m3,m4,m5,m6,m7,m8a,m8,m9,m10,m11,MRW:2004,m12,m13,m14}.
The measures $\Delta$ and $\Sigma$ have several advantages:
they are dimensionless and give
a common scale required for a quantitative comparison of the
e-by-e fluctuations (see more details in Ref.~\cite{GGP:2013}).

The strongly intensive measures $\Delta$ and $\Sigma$
are defined using two extensive quantities, i.e. the quantities
proportional to the system volume.
A popular example of such a pair of extensive
variables is: the transverse momentum $P_T=p_T^{(1)}+\dots
p_T^{(N)}$, where $p^{(i)}_T$ is the absolute value of the
$i^{{\rm th}}$ particle transverse momentum, and the number of
particles $N$.
The measures $\Delta[P_T,N]$ and $\Sigma[P_T,N]$ were
studied recently within the
UrQMD simulations in Ref.~\cite{KG:APP2012,GGP:2013} and within the
GCE
formulation for the ideal Bose and Fermi gases in Ref.~\cite{GR:2013}.
The basic
properties of the $\Delta[P_T, N]$ and $\Sigma[P_T, N]$ measures
were also tested using the Monte Carlo simulations and
analytical models in Ref.~\cite{GG:2014}.

The measures
$\Delta[K,\pi]$ and $\Sigma[K,\pi]$, in the case of
multiplicities of charged kaons, $K=K^++K^-$, and
pions, $\pi=\pi^++\pi^-$, were considered within the
hadron-string dynamics transport model in Ref.~\cite{HSD}.
Note that the NA49 and NA61/SHINE
Collaborations have  already started to use the strongly intensive measures
$\Delta$ and $\Sigma$ for the studies of e-by-e fluctuations in $p+p$ and A+A
collisions (see Ref.~\cite{GGS} and references therein).

%
%
\vspace{0.3cm}
In the present paper the strongly intensive measures for
particle number fluctuations are studied  \cite{GG:2011}:
 \eq{\label{Delta-12}
 &\Delta[N_1,N_2]
 ~=~ \frac{1}{C_{\Delta}} \Big[ \langle N_2\rangle\,
      \omega[N_1] ~-~\langle N_1\rangle\, \omega[N_2] \Big]~, \\
&\Sigma[N_1,N_2]
 ~=~ \frac{1}{C_{\Sigma}} \Big[ \langle N_1\rangle\,
      \omega[N_2] ~+~\langle N_2\rangle\, \omega[N_1] ~-~2\Big(\langle N_1\,N_2\rangle
      ~-~\langle N_1\rangle \langle N_2\rangle\Big)\Big]~,\label{Sigma-12}
}
where  $N_1$ and $N_2$
are the multiplicities for hadrons of types 1 and 2,
\eq{\label{omegaN1N2}
\omega[N_1]~=~\frac{\langle N_1^2 \rangle~ -~ \langle N_1 \rangle^2}{
\langle N_1 \rangle}~,~~~~\omega[N_2]~=~\frac{\langle N_2^2 \rangle~ -~ \langle N_2 \rangle^2}{
\langle N_2 \rangle}~
}
are the scaled variances of  $N_1$ and $N_2$ distributions, and
$C_{\Delta}$ and $C_{\Sigma}$ are the normalization factors.
A notation $\langle \dots \rangle$ represents the e-by-e averaging.

%
%
In a classical thermal system of non-interacting
particles within the GCE formulation, the partition function $Z$ for a mixture
of particles
of types 1 and 2 is equal to
\eq{\label{Z12}
Z~=~\sum_{N_1=0}^{\infty}\sum_{N_2=0}^{\infty}\frac{z_1^{N_1}}{N_1!}~\frac{z_2^{N_2}}{N_2!}~,
}
where $z_i$ ($i=1,2$) is the so called one-particle partition function,
\eq{\label{zi}
z_i~=~V\,\exp\left(\frac{\mu_i}{T}\right)\,\frac{d_im_i^2T}{2\pi^2}\,K_2\left(\frac{m_i}{T}\right)~.
}
In Eq.~(\ref{zi}), $d_i$, $m_i$, and $\mu_i$ are, respectively,
the degeneracy factor, mass, and chemical potential for particles of the $i$th type, whereas
$V$ and $T$ are the system volume and temperature, respectively.  According to Eq.~(\ref{Z12})
a joint probability distribution ${\cal P}(N_1,N_2)$ for variables $N_1$ and $N_2$ is just a
simple product of two Poissonian distributions
\eq{\label{P12}
{\cal P}(N_1,N_2)~=~\frac{1}{Z}\,\frac{z_1^{N_1}}{N_1!}\,\frac{z_2^{N_2}}{N_2!}~=~P(N_1)\,P(N_2)~,
}
with
\eq{\label{Pi}
P(N_i)~=~\frac{\langle N_i\rangle^{N_i}}{N_i!}\,\exp\left(-\,\langle N_i\rangle\right)~.
}
Taking e-by-e averaging with a probability distribution ${\cal P}(N_1,N_2)$ (\ref{P12}),
one finds
\eq{\label{omIPM}
& \omega[N_1]~=~\omega[N_2]~=~1~,\\
& \langle N_1N_2\rangle~-~\langle N_1\rangle \langle N_2\rangle~=~0~.\label{corIPM}
}
According to Eq.~(\ref{corIPM}) there are no correlations between $N_1$ and $N_2$ numbers
in the ideal Boltzmann gas within the GCE. 

In Ref.~\cite{GGP:2013} the special normalization has been
proposed for the $\Delta$ and $\Sigma$ fluctuation measures. It is
used in the present study,
and for the case under consideration it reads:
\eq{\label{CDCS}
C_{\Delta}~=~\langle N_2\rangle~-~\langle N_1\rangle~,~~~~~C_{\Sigma}~=~\langle N_1\rangle~
+~\langle N_2\rangle~.
}
The multi-component ideal Boltzmann gas before decays of resonances will be denoted
as IB-GCE. It gives an important
example of independent particle model (see Ref.~\cite{GGP:2013}). As shown above
this model
satisfies Eqs.~(\ref{omIPM}) and (\ref{corIPM}).
The special choice of normalization factors (\ref{CDCS})
leads then to
%
\eq{\label{DS-IPM}
\Delta[N_1,N_2]~=~\Sigma[N_1,N_2]~=~1~.
}
The independent particle model, together with the IB-GCE,
plays an important role as the {\it reference
model}. The deviations of real data from its
results (\ref{DS-IPM}) can be used to clarify the physical
properties of the system, i.e., relation (\ref{DS-IPM}) provides
a common scale required for a quantitative comparison of the
e-by-e fluctuations.
Note also that with normalization factors (\ref{CDCS}),
both $\Delta[N_1,N_2]$ and $\Sigma[N_1,N_2]$
become symmetric:
\eq{\label{sym}
\Delta[N_1,N_2]~=~\Delta[N_2,N_1]~,~~~~~\Sigma[N_1,N_2]~=~\Sigma[N_2,N_1]~.
}

Effects of quantum statistics change the results
(\ref{omIPM}).
Bose  (Fermi) statistics leads to
$\omega[N]$ larger (smaller) than unity. However, for the hadron systems
created in A+A collisions the corrections due to quantum statistics
are small. The largest effects are for pions  at high
temperature, when $\omega[N]\cong 1.1$,
and negligible for other hadrons and resonances (see, e.g., Ref.~\cite{HGM}).
Therefore, relations (\ref{DS-IPM}) remain approximately valid for
the quantum gases.

A presence of resonances decaying into particle species 1 and/or 2
change the results (\ref{omIPM}) and (\ref{corIPM}). Thus,
relations (\ref{DS-IPM}) are also changed.
These effects of particle number fluctuations and correlations due to decays of
resonances will be a subject of our study.

Before a discussion of the effects of resonances
we remind shortly the role of volume fluctuations within a thermal model.
The average multiplicities are then assumed to be proportional to the
system volume $V$,
\eq{\label{NV}
 \langle N_1\rangle~ =~ \rho_1\, \langle V \rangle~,~~~~
\langle N_2\rangle~ = ~\rho_2\, \langle V \rangle~,
}
where $\rho_1$ and $\rho_2$ are the corresponding particle densities
assumed to be independent of $V$.
The scaled variances and correlations can be then presented as \cite{GG:2011}:
\eq{\label{om1V}
& \omega[N_1]~=~\frac{\langle N_1^2\rangle_V~-~\langle N_1\rangle_V ^2}{\langle N_1\rangle_V}~
+~\rho_1\,\omega[V]~,\\
&\omega[N_2]~=~\frac{\langle N_2^2\rangle_V~-~\langle N_2\rangle_V^2}{\langle N_2\rangle_V}~
+~\rho_2\,\omega[V]~, \label{om2V} \\
& \langle N_1N_2 \rangle ~-\langle N_1\rangle \langle N_2\rangle~=~\langle N_1N_2\rangle_V
~-~\langle N_1\rangle_V~\langle N_2\rangle_V~+~\rho_1 \rho_2 \, \langle V\rangle \,\omega[V]~,\label{12V}
}
where $\langle \ldots\rangle_V$ denotes the averaging at fixed volume $V$,
 and $\omega[V]\equiv (\langle V^2\rangle -\langle V\rangle^2)/\langle V\rangle$
describes the volume e-by-e fluctuations.
The scaled variances (\ref{om1V},\ref{om2V}) and correlation term
(\ref{12V}) have additional contributions proportional to $\omega[V]$. As already
mentioned, the
volume fluctuations are usually rather large. Besides, it is rather difficult
to control them experimentally.
Thus, it is not easy to extract physical information
from quantities (\ref{om1V}-\ref{12V}).
On the other hand, substituting Eqs.~(\ref{NV}-\ref{12V}) into
Eqs.~(\ref{Delta-12},\ref{Sigma-12})
one finds that all terms
proportional to $\omega[V]$ are canceled out, i.e. $\Delta[N_1,N_2]$
and $\Sigma[N_1,N_2]$ are independent of the average
volume and volume fluctuations for the statistical
systems within the GCE.

\section{Analytical Examples}\label{sec-AM}

Resonance decay is a probabilistic process.
Introducing probabilities $b_r^R$ for $r$-th  decay channel of $R$-th resonance and numbers
$n_{i,r}^R$ of $i$-th final particles  produced in these decay channels, one finds
for the average multiplicity and scaled variance of $i$-th type of hadrons
from decays of $R$-th type of resonances \cite{HGM}:
\eq{\label{resN}
&\langle N_i\rangle
~=
~\langle n_{i}\rangle_R ~\langle R\rangle~ ,\\
&\omega[N_i]~=~\frac{\sum_r b_r^R\,\left(n_{i,r}^R\right)^2~-~\left(\sum_r b_r^R\,n_{i,r}^R\right)^2}
{\langle n_i\rangle_R}~+~ \langle n_i\rangle_R \,\omega[R]~\equiv~\omega^*[N_i]
~+~ \langle n_i\rangle_R \,\omega[R]~,\label{resN2}
}
where $\langle R\rangle $ denotes the average number of $R$-th resonances,
and $\langle n_i\rangle_R \equiv \sum_r b_r^R\,n_{i,r}$ means the average number of $i$-th hadrons
produced from the decay of one $R$-th resonance.
Note that different decay channels $r$ in Eqs.~(\ref{resN}) and (\ref{resN2})
are defined in a way that final state with only stable (with respect to strong and electromagnetic decays)
hadrons are counted, and probabilities $b_r^R$ satisfy the normalization condition, $\sum_r b_r^R=1$.
Resonances act as independent sources of particles:
the first term $\omega^*[N_i]$ in the right hand side of Eq.~(\ref{resN2}) describes
the $N_i$ fluctuations
from a single source, and the second term appears
due to the fluctuations of the number of sources.

Some examples are appropriate to illustrate Eq.~(\ref{resN2}). Let us assume
a presence of two types of decay channels $1$ and $2$ with $n_{i,1}^R=1$ and
$n_{i,2}^R=0$, and with the corresponding probabilities $b_1^R=p$ and $b_2^R=1-p$.
It then follows from Eq.~(\ref{resN2})
\eq{\label{1}
\omega[N_i]~=~1~-~p~+~p\,\omega[R]~.
}
In the GCE formulation for the hadron-resonance gas, one finds $\omega[R]\cong 1$
as the effects of quantum statistics are negligible for resonances. From Eq.~(\ref{1}),
one then obtains $\omega[N_i]\cong 1$, thus, resonance decays do not change
Eq.~(\ref{omIPM}). At $p\ll 1$ the main contribution to $\omega[N_i]$ in
Eq.~(\ref{1}) comes from $\omega^*[N_i]=1-p\cong 1$, while at $p\cong 1$ from $p\,\omega[R]\cong 1$.
However, resonance contributions to $\omega[N_i]$ becomes really important
if there are decay channels with two particles of $i$-th type. Let us assume
again two types of decay channels $1$ and $2$  with probabilities $b_1^R=p$ and $b_2^R=1-p$,
for which $n_{i,1}^R=2$ and $n_{i,2}^R=0$. One then finds from Eq.~(\ref{resN2})
\eq{\label{2}
\omega[N_i]~=~2\,(1~-~p)~+~2\,p\,\omega[R]~\cong~2~,
}
i.e., the scaled variance $\omega[N_i]$ is increased by a factor of 2.
Therefore, a presence of resonances decaying to two (or more) particles of $i$-th type
can enlarge $\omega[N_i]$ essentially.

If resonance $R$ has decay channels where particles of types 1 and 2 appears
simultaneously in the final state,
the correlation between numbers $N_1$ and $N_2$ appears:
\eq{
\langle N_1N_2\rangle~ -~\langle N_1\rangle \langle N_2\rangle~&=~\Big(\langle R^2\rangle ~
-~\langle R\rangle^2\Big) \langle n_1\rangle_R \langle n_2\rangle_R~\nonumber \\
&+~ \langle R\rangle
\Big(\langle n_1n_2 \rangle_R~-~ \langle n_1\rangle_R\langle n_2\rangle_R\Big)
~ \cong ~\langle R\rangle~\langle n_1n_2 \rangle_R~, \label{cor}
}
where $\langle n_1n_2 \rangle_R\equiv \sum_r b^R_r n_{1,r}^Rn_{2,r}^R$. Thus,
Eq.~(\ref{corIPM}) is no more valid.

\vspace{0.3cm}
Our goal in this section will be to calculate the
strongly intensive measures (\ref{Delta-12},\ref{Sigma-12})
for two simple analytical examples. The system of non-interacting
Boltzmann particles and resonances within the GCE will be considered.

In the first example, resonances
decaying into two positively charged particles are considered.
The prominent example is the decay of $\Delta^{++}$-resonance,
$\Delta^{++}\rightarrow p + \pi^+$.
Note that the systems with
positive net baryon number (and positive electric charge)
are created in A+A or $p+p$. Thus,
an effect of resonance decays into two negatively charged particles
is much weaker
and can be safely neglected. This is, however, not the case for RHIC and LHC energies,
where the baryonic and electric charge densities are very small.
Both processes -- resonance decays into two positively charged and two negatively charged
hadrons -- become then comparable.

In the second example, the correlated pairs of charged pions coming from  the decays of resonances
are considered. The main source of these $\pi^+\pi^-$-pairs  are meson resonances, e.g.,
$\rho^0,~\omega \rightarrow \pi^++\pi^-$.

For simplicity, in both model examples  an existence of only one type
of resonances decaying with probability 1 into two positively charged hadrons
or into $\pi^+\pi^-$-pair is assumed. These resonances will
be denoted as $R_{++}$ and $R_{\pi\pi}$, respectively. The same notations will be used
for their multiplicities.

\subsection{Resonance Decays to Two Positively Charged Hadrons}\label{ss-pm}
Resonance decays into two
final particles of the same type 1 (or 2) lead to positive contributions
to the corresponding scaled variance $\omega[N_1]$ (or $\omega[N_2]$).
In our first model example, we consider $N_1=N_-$ and $N_2=N_+$, where
$N_-$ and $N_+$ are the numbers of negatively and positively charged hadrons,
respectively.
We assume,
%
\eq{\label{NpR}
%
N_-~=~n_-~,~~~~~~~
N_+~=~n_+~+~2R_{++}~,
}
where $R_{++}$ is the number of resonances which decay
into two positively charged hadrons, and the numbers of negatively and positively
charged hadrons from other sources are denoted as
$n_-$ and $n_+$, respectively.
In the GCE, the numbers  $n_+$ and $R_{++}$ are not correlated. Therefore,
the particle number distribution $W(N_+)$ is equal to:
\eq{
W(N_+)~=~\sum_{n_+,R_{++}}P_+(n_+) \, P_{R}(R_{++})\,\delta(N_+ -n_+-2R_{++})~.
%
}
%

The first and second moments of $P_-(N_-)$, $P_+(n_+)$, and $P_R(R_{++})$ distributions
are denoted as ($k=1,\,2$)
\eq{\label{moments}
\langle N_-^k\rangle=\sum_{N_-} P_-(N_-)\,N_-^k~,~~~~~
\langle n_+^k\rangle=\sum_{n_+} P_+(n_+)\,n_+^k~,~~~~~
\langle R^k\rangle=\sum_{R_{++}} P_{R}(R_{++})\,R_{++}^k~.
}
The first and second moments of the  $N_+$ distribution are calculated  as
\eq{\label{N+}
%
%
\langle N_+\rangle~=~ \langle n_+\rangle ~+~2\langle R_{++}\rangle~, ~~~~
\langle N_+^2\rangle~=~\langle n_+^2\rangle ~+~4\langle n_+\rangle ~\langle R_{++}\rangle~+~4\langle R_{++}^2\rangle~.
}
The scaled variance of the $N_+$ distribution then equals
\eq{\label{omegaN+}
%
%
\omega[N_+]~
\equiv~\frac{\langle N_+^2\rangle ~-~\langle N_+ \rangle^2}{\langle N_+ \rangle}~
=~
\frac{\langle n_+\rangle\,\omega[n_+]+
4\langle R_{++} \rangle\,\omega[R_{++}]}{\langle N_+\rangle }~,
}
and for the $\Delta$ measure (\ref{Delta-12}) one finds\footnote{We do not
make any assumptions about the correlations between $N_+$ and $N_-$ numbers.
Thus, we do not attempt here to calculate the $\Sigma[N_-,N_+]$ measure.
An example of $\Sigma[\pi^+,\pi^-]$ calculations is considered in the next subsection. }:
\eq{
\Delta[N_-,N_+]~=~\frac{1}{\langle N_+\rangle -\langle N_-\rangle}\Big[\langle N_+\rangle
\,\omega[N_-]~-~
\langle N_-\rangle ~ \frac{\langle n_+\rangle \omega[n_+]+4\langle R_{++}\rangle \omega[R_{++}]}{\langle N_+\rangle }\Big]~.
}
Within approximations,
\eq{\label{appr}
\omega[n_+]~\cong \omega[N_-]~,~~~~~~~\omega[R_{++}]~\cong~1~,
%
}
one obtains:
\eq{\label{DR}
\Delta[N_-,N_+]
~ \cong~
\omega[N_-]~-~\frac{2\langle N_-\rangle \,\langle R_{++}\rangle\,(2~-~\omega[N_-]\,)}{\langle N_+\rangle
\,[\,\langle N_+\rangle-\langle N_-\rangle\,]}~.
%
}
%
From Eq.~(\ref{DR}) it follows
\eq{\label{Rp}
\frac{\langle R_{++}\rangle}{\langle N_+\rangle }~\cong~\frac{\langle N_+\rangle -\langle N_-\rangle}{2\,
\langle N_-\rangle\,(2-\omega[N_-])}~\Big[\,\omega[N_-]~-~
\Delta[N_-,N_+]\,\Big]~.
}
Note that within approximation (\ref{appr}), one can also calculate $\langle R_{++}\rangle$ from
Eqs.~(\ref{N+}) and (\ref{omegaN+}):
\eq{\label{Rpp}
\frac{\langle R_{++}\rangle}{\langle N_+\rangle}~\cong~\frac{\omega[N_+]~-~\omega[N_-]}{2\,(2-\omega[N_-]\,)}~,
}
which is identical to expression (\ref{Rp}).

For $\omega[N_-]\cong 1$, Eq.~(\ref{Rp})
is further simplified to
\eq{\label{Rp1}
 \frac{\langle R_{++}\rangle}{\langle N_+\rangle }~\cong~\frac{\langle N_+\rangle -\langle N_-\rangle}{2\,
\langle N_-\rangle}~\Big[\,1~-~\Delta[N_-,N_+]\,\Big]~.
%
}

\subsection{Resonance Decays to ${\bf \pi^+\pi^-}$ Pairs }\label{ss-pi}
Resonance decays, when particle species 1 and 2 appear simultaneously
among the decay products, lead to the (positive) correlations between $N_1$ and $N_2$
numbers, i.e. the left hand side of Eq.~(\ref{corIPM}) become positive.

In our second model example,
$N_1=\pi^+$ and $N_2=\pi^-$ are the  multiplicities
of positively and negatively charged  pions, respectively.
A presence of two components
is assumed: the correlated pion pairs coming from decays,
$R_{\pi\pi}\rightarrow \pi^++\pi^-$,
and the uncorrelated $\pi^+$ and $\pi^-$
from other sources. The $\pi^+$ and $\pi^-$ numbers are then equal to:
%
\eq{
\pi^+~=~n_+~+~R_{\pi\pi}~,~~~~~~\pi^-~=~n_-~+~R_{\pi\pi}~,
}
where $R_{\pi\pi}$ is the number of resonances decaying
into $\pi^+\pi^-$ pairs, while $n_+$ and $n_-$ are the numbers
of uncorrelated
$\pi^+$ and $\pi^-$, respectively. The number distribution of $\pi^+$ and $\pi^-$ is
\eq{
W(\pi^+,\pi^-)=\sum_{n_+,n_-,R_{\pi\pi}} P_+(n_+)\, P_-(n_-)
\, P_{R}(R_{\pi\pi})\,\delta(\pi^+ -n_+-R_{\pi\pi})
\,\delta(\pi^--n_--R_{\pi\pi}).
}
%

%
%
%
%

The first and second moments of $\pi^+$ and  $\pi^-$ are calculated  as
\eq{
& \langle \pi^+\rangle ~=~ \langle n_+\rangle ~+~\langle R_{\pi\pi}\rangle~,~~~~~
\langle (\pi^+)^2\rangle ~=~\langle n_+^2\rangle ~+~2\langle n_+\rangle~\langle R_{\pi\pi}\rangle~+~\langle
R_{\pi\pi}^2\rangle ~,\\
& \langle \pi^-\rangle ~=~ \langle n_-\rangle~+~\langle R_{\pi\pi}\rangle ~,~~~~~
\langle (\pi^-)^2\rangle ~=~\langle n_-^2\rangle~+~2\langle n_-\rangle ~\langle R_{\pi\pi}
\rangle~+~\langle R_{\pi\pi}^2\rangle~,\\
& \langle \pi^+\pi^-\rangle ~=~\langle n_+\rangle ~ \langle n_-\rangle~+~\langle n_+\rangle~
\langle R_{\pi\pi}\rangle ~ +~
\langle n_-\rangle ~\langle R_{\pi\pi}\rangle~+~\langle R_{\pi\pi}^2\rangle~,
}
where $\langle n_+^k\rangle $, $\langle n_-^k\rangle $, and $\langle R_{\pi\pi}^k\rangle$ are similar
to those in Eq.~(\ref{moments}).
For the
correlation term one obtains
\eq{
%
%
%
\rho[\pi^+,\pi^-]~\equiv~\frac{ \langle \pi^+\pi^-\rangle ~-~\langle \pi^+\rangle ~\langle \pi^-\rangle}
{\langle \pi^+\rangle ~+~\langle \pi^-\rangle } ~
=~\frac{\langle R_{\pi\pi}^2\rangle -\langle R_{\pi\pi}\rangle^2}
{\langle \pi^+\rangle ~+~\langle \pi^-\rangle }
~\equiv~
\frac{\langle R_{\pi\pi}\rangle ~ \omega[R_{\pi\pi}]}
{\langle \pi^+\rangle ~+~\langle \pi^-\rangle }~.\label{pi-corr}
}
For $\Delta$ and $\Sigma$ measures one finds
\eq{
&\Delta[\pi^+,\pi^-]~\equiv~\frac{1}{\langle \pi^- \rangle-\langle \pi^+ \rangle}
~\Big[\langle \pi^- \rangle\,\omega[\pi^+]~-~
\langle \pi^+ \rangle \, \omega[\pi^-]\Big]~,\\
&\Sigma[\pi^+,\pi^-]~=~\frac{1}{\langle \pi^- \rangle+\langle \pi^+ \rangle}
~\Big[\langle \pi^- \rangle\,\omega[\pi^+]~+~
\langle \pi^+ \rangle \, \omega[\pi^-]
~ -~2\langle R_{\pi\pi} \rangle\omega[R_{\pi\pi}]\Big ]~.
 }
Using approximate relations,
\eq{\label{omega-pm}
%
\omega[\pi^+]~\cong ~\omega[\pi^-]~\equiv~\omega^*,~~~~
\omega[R_{\pi\pi}] \cong 1~,~
}
one obtains
\eq{\label{Dpm}
& \Delta[\pi^+,\pi^-]~\cong~
\omega^*~,
\\
& \Sigma[\pi^+,\pi^-]
~\cong~ \omega^*~-~\frac{2\langle R_{\pi\pi} \rangle}{\langle \pi^-\rangle +\langle \pi^+ \rangle}~.
\label{Spm}
}
From Eqs.~(\ref{pi-corr}) and
(\ref{Spm}) it follows, respectively,
\eq{
\label{R0}
& \frac{\langle R_{\pi\pi} \rangle}{\langle \pi^-\rangle +\langle \pi^+ \rangle}~\cong~\rho[\pi^+,\pi^-]~,\\
&\frac{\langle R_{\pi\pi} \rangle}{\langle \pi^-\rangle +\langle \pi^+ \rangle}~\cong~
\frac{\omega^* ~-~ \Sigma[\pi^+,\pi^-]}{2}
~,\label{R00}
}
i.e. the average number of $R_{\pi\pi}$-resonances can be calculated using the
measurable quantities.
For $\omega^*\cong 1$, Eq.~(\ref{Spm}) is further simplified to
\eq{\label{R}
\frac{\langle R_{\pi\pi} \rangle}{\langle \pi^-\rangle
+\langle \pi^+ \rangle}~\cong~\frac{1~-~\Sigma[\pi^+,\pi^-]}{2}~.
}

Equation (\ref{Dpm}) does not give any information on the number
of resonances. Besides,
for a physically interesting case $\langle \pi^- \rangle\cong \langle \pi^+ \rangle$,
there is an uncertainty, $0/0$, in $\Delta[\pi^+\pi^-]$.
This may lead to numerical problems in using the $\Delta[\pi^+,\pi^-]$
measure in data analysis.
%
%
%
%

\section{Hadron-Resonance Gas Model}\label{sec-HG}
The thermal hadron-resonance gas model (HGM) within
the GCE formulation is used in this section to calculate the $\Delta$
and $\Sigma$ measures considered in Secs.~\ref{ss-pm} and \ref{ss-pi}.
We use the THERMUS package \cite{THERMUS}
which includes particles and resonances (mesons up to $K_4^*(2045)$ and baryons up to $\Omega^-$),
quantum statistics, as well as the  widths of resonances.

The results of the HGM in the GCE are presented in Tables \ref{table1} and \ref{table2}
for central Pb+Pb (or Au+Au) collisions at different center of mass energy
per nucleon pair $\sqrt{s_{NN}}$.
The temperature $T$
and baryonic chemical potential $\mu_B$ at different collision energies
were found by fitting the hadron multiplicities. For central heavy-ion collisions
they can be
parameterized  as \cite{Tmu}:
\eq{\label{T}
T~&=~0.166~{\rm GeV}~-~0.139~{\rm GeV}^{-1}\mu_B^2~-~0.053~{\rm GeV}^{-3}\mu_B^4~,\\
\mu_B~&= ~\frac{1.308~{\rm GeV}}{1~+~0.273~{\rm GeV}^{-1}\sqrt{s_{NN}}}~.\label{mu}
}
The strangeness suppression factor $\gamma_S$ \cite{Raf}, which regulates incomplete
strangeness equilibration, is taken as \cite{gammaS}:
\eq{
\gamma_S~=~1~-~0.396\, \exp(-1.23\,T/\mu_B)~.
}
The details of calculations
of the first and second moments of particle number distributions in the HGM
can be found in Ref.~\cite{HGM}.%
\begin{table}
    \begin{center}
        \begin{tabular}{||c||c|c||c|c|c|c||c||c|c||} \hline\hline
     $\sqrt{s_{NN}} $
   &\ $T$
   &\ $\mu_B$
   &\ $\langle N_+ \rangle/\langle N_- \rangle$
   &\ $\omega[N_+]$
   &\ $\omega[N_-]$
   &\ $\Delta[N_-,N_+]$
   & \multicolumn{3}{c||}{$\langle R_{++} \rangle/\langle N_+ \rangle$}
\\ \hline
   \ [GeV]         & \ [MeV]  &[MeV]  &\ GCE  &\ GCE   &\ GCE   &\ GCE  &\ Eq.(\ref{R-HGM1})&\ Eq.(\ref{Rp})
   &\ Eq.(\ref{Rp1})  \\
                 \hline \hline
             \ 6.27 &\ 130.8 &\ 482.4 &\ 1.624 &\ 1.237 &\ 1.071  &\ 0.807 &\ 0.11 &\ 0.09  &\ 0.06    \\\hline
             \ 7.62 &\ 139.2 &\ 424.6 &\ 1.500 &\ 1.232 &\ 1.080  &\ 0.778 &\ 0.11 &\ 0.08  &\ 0.06    \\\hline
             \ 8.77 &\ 144.2 &\ 385.4 &\ 1.429 &\ 1.226 &\ 1.087  &\ 0.763 &\ 0.10 &\ 0.08  &\ 0.05    \\\hline
             \ 12.3 &\ 153.0 &\ 300.2 &\ 1.301 &\ 1.210 &\ 1.101  &\ 0.740 &\ 0.09  &\ 0.06  &\ 0.04    \\\hline
             \ 17.3 &\ 158.6 &\ 228.6 &\ 1.213 &\ 1.195 &\ 1.114  &\ 0.730  &\ 0.08 &\ 0.05  &\ 0.03    \\\hline
             \hline
        \end{tabular}
         \caption{The results of the GCE HGM
         for central collisions of heavy ions
         at different center of mass  energy per nucleon pair. At given $\sqrt{s_{NN}}$
         the temperature $T$ and baryonic chemical potential $\mu_B$
         are calculated according to Eq.~(\ref{T}) and (\ref{mu}), respectively.
           } \label{table1}
            \end{center}
\end{table}
\begin{table}
    \begin{center}
        \begin{tabular}{||c||c|c||c|c|c|c||c||c|c||} \hline\hline
   \ $\sqrt{s_{NN}} $ &\ $T$ &\ $\mu_B$ &\ ~$\rho[\pi^+,\pi^-]$~
    &\ ~$\omega[\pi^+]$~ &\ ~$\omega[\pi^-]$~
   &\  ~$\Sigma[\pi^+,\pi^-]$~
    &  \multicolumn{3}{c||}{$\langle R_{\pi\pi}\rangle/(\langle \pi^-\rangle + \langle \pi^+\rangle)$}  \\ \hline
   \ [GeV] & \ [MeV]  &[MeV]  &\ GCE  &\ GCE  &\ GCE  &\ GCE  &\ Eq.(\ref{R-HGM})&\ Eq.(\ref{R00}) &\ Eq.~(\ref{R})   \\
                 \hline \hline
             \ 6.27 &\ 130.8 &\ 482.4 &\ 0.132  &\ 1.063 &\ 1.074   &\ 0.804 &\ 0.11 &\ 0.14    &\ 0.10   \\\hline
             \ 7.62 &\ 139.2 &\ 424.6 &\ 0.155  &\ 1.072 &\ 1.083   &\ 0.768 &\ 0.13 &\ 0.16    &\ 0.12   \\\hline
             \ 8.77 &\ 144.2 &\ 385.4 &\ 0.170  &\ 1.079 &\ 1.089   &\ 0.744 &\ 0.14 &\ 0.17    &\ 0.13   \\\hline
             \ 12.3 &\ 153.0 &\ 300.2 &\ 0.199  &\ 1.092 &\ 1.101   &\ 0.699 &\ 0.15 &\ 0.20    &\ 0.15   \\\hline
             \ 17.3 &\ 158.6 &\ 228.6 &\ 0.219  &\ 1.102 &\ 1.109   &\ 0.668 &\ 0.16 &\ 0.22    &\ 0.17   \\\hline
             \ 200      &\ 165.9 &\ 23.5  &\ 0.246 &\ 1.116 &\ 1.117   &\ 0.624 &\ 0.17 &\ 0.25 &\ 0.19   \\\hline
             \ 5500     &\ 166.0 &\ 0.87  &\ 0.246 &\ 1.117 &\ 1.117   &\ 0.624 &\ 0.17 &\ 0.25 &\ 0.19   \\\hline
             \hline
        \end{tabular}
         \caption{The results of the GCE HGM at different center of mass  energy per nucleon pair
         for central collisions of heavy ions. At given $\sqrt{s_{NN}}$
         the temperature $T$ and baryonic chemical potential $\mu_B$
         are calculated according to Eq.~(\ref{T}) and (\ref{mu}), respectively.
                    } \label{table2}
            \end{center}
\end{table}

The results of the HGM in the GCE
for the ratio of average multiplicities $\langle N_+\rangle/\langle N_-\rangle$,
scaled variances $\omega[N_+]$ and $\omega[N_-]$, and strongly intensive measure
$\Delta[N_-,N_+]$ are presented in Table~I. The values of $\sqrt{s_{NN}}$ in Table~I
correspond to the projectile momenta $p_{\rm lab}=20A$, $30A$, $40A$, $80A$, and $158A$~GeV/c.
Note that in the IB-GCE introduced in Sec.~\ref{SI} one has $\omega[N_+]=\omega[N_-]=1$
and $\Delta[N_-,N_+]=1$.
Higher values of $\omega[N_+]$ than those of $\omega[N_-]$, presented in Table \ref{table1},
are just a consequence of the contribution due to
the $R_{++}$ decays.

The GCE HGM values 
from middle columns of Table I can be used to estimate
$\langle R_{++} \rangle/\langle N_+\rangle$ according to Eqs.~(\ref{Rp}) and (\ref{Rp1}).
The value of $\langle R_{++}\rangle/\langle N_+\rangle$ can be
also straightforwardly calculated in the HRG model as
\eq{\label{R-HGM1}
\frac{\langle R_{++}\rangle}{\langle N_+\rangle} ~=~ \frac{\sum_{R} \langle R\rangle \,b^R_{++}}{\langle N_+\rangle}~,
}
where $b^R_{++}$ is the probability of resonance $R$
to decay with two positively charged hadrons among its decay products, and the sum in Eq.~(\ref{R-HGM1})
is taken over all types of resonances.
The results obtained from Eqs.~(\ref{Rp}), (\ref{Rp1}), and (\ref{R-HGM1}) are presented in Table~I.
%
Two positively charged hadrons produced by resonance decays are most likely $p$ and $\pi^+$,
and the $\Delta^{++}$ resonance gives a dominant contribution to the sum in Eq.~(\ref{R-HGM1}).
For example, at $\sqrt{s_{NN}}=6.27$~GeV, the lowest state $\Delta^{++}(1232)$ gives about 48\%,
and all  $\Delta^{++}$ states about 75\%, of the whole sum in Eq.~(\ref{R-HGM1}).


%
%
%
%

The GCE HGM values for $\rho[\pi^+,\pi^-]$, $\omega[\pi^+]$, $\omega[\pi^-]$, and $\Sigma[\pi^+,\pi^-]$
are presented in
Table~II.
One can use
Eqs.~(\ref{R00}) and (\ref{R})
to estimate the value of $\langle R_{\pi\pi}\rangle/(\langle\pi^-\rangle
+\langle \pi^+\rangle)$.
The values of $\rho[\pi^+,\pi^-]$ in Table II demonstrate that  Eqs.~(\ref{R0}) and (\ref{R00})
lead to almost identical results for $\langle R_{\pi\pi}\rangle/(\langle \pi^-\rangle +\langle \pi^+\rangle)$.
This is because our
assumption (\ref{omega-pm}) is rather accurately fulfilled in the HGM.
Similar to Eq.~(\ref{R-HGM1}) the HGM value of $\langle R_{\pi\pi}\rangle/(\langle\pi^-\rangle
+\langle \pi^+\rangle)$ can be straightforwardly calculated
as
%
\eq{\label{R-HGM}
\frac{\langle R_{\pi\pi}\rangle}{\langle N_+\rangle} ~=~ \frac{\sum_{R} \langle R\rangle \,b^R_{\pi\pi}}
{\langle N_+\rangle}~,\
%
}
where $b^R_{\pi\pi}$ is the probability of resonance $R$
to decay with $\pi^+$ and $\pi^-$ among its decay products.
The results calculated from Eqs.~(\ref{R00}),  (\ref{R}), and (\ref{R-HGM})
are presented in Table II. The sum in Eq.~(\ref{R-HGM})
includes the contributions from numerous
mesonic and baryonic resonances.
However,
$\rho^0(770)$ and $\omega(782)$
give the dominant contribution, e.g.,  about 50\% at $\sqrt{s_{NN}}=17.3$~GeV.

In general, the results presented in Tables I and II are in a qualitative
agreement with  the analytical results of Sec.~\ref{sec-AM}.


\section{UrQMD Simulations}\label{sec-UrQMD}
In this section, the UrQMD \cite{urqmd} model, i.e. the relativistic transport
approach to A+A collisions, is used. We consider $\pi^+$ and $\pi^-$ fluctuations and correlations
discussed in Sec.~\ref{ss-pi} to illustrate the role of centrality selection,
limited acceptance, and global charge conservation in A+A collisions.
The samples of
5\% central Pb+Pb collision events at $\sqrt{s_{NN}}= 6.27$ and 17.3~GeV are considered.
These results will be compared with UrQMD simulations in the
most central Pb+Pb collisions at zero impact parameter, $b=0$~fm,
and in  $p+p$ reactions at the same collision energies.
Several mid-rapidity windows
$-\Delta y/2 <y<\Delta y/2$ for final  $\pi^+$ and $\pi^-$ particles
are considered.

A width of the rapidity window $\Delta y$ is an important parameter.
The two hadrons which are the products of a resonance decay have, in average, a
rapidity difference of the order of unity. Therefore, while searching for the effects of resonance
decays one should choose $\Delta y\geq 1$
to enlarge a probability for simultaneous hit into the rapidity window $\Delta y$
of both  correlated hadrons  (e.g., $\pi^+$ and $\pi^-$) from resonance decays.
Thus, $\Delta y$ should be {\it large} enough.  However, $\Delta y$ should be
{\it small} in  comparison to the whole rapidity interval $\Delta Y \approx \ln(\sqrt{s_{NN}}/m)$
accessible for final hadron with mass $m$.
 Only for  $\Delta y\ll \Delta Y$ one can expect a validity of
the GCE results presented in Secs.~\ref{sec-AM} and \ref{sec-HG}.
Considering a small part of the statistical system, one does not need to impose
the restrictions of the exact global charge conservations: the GCE which only regulates
the average values of the conserved charges  is fully acceptable.
For large $\Delta y$, when the detected hadrons correspond to an essential
part of the whole system, the effects of the global charge conservation become more important.
In the HGM this should be treated within the CE, where the conserved charges
are fixed for all microscopic states. The global charge conservation influences
the particle number fluctuations  and introduces additional
correlations between numbers of different particle species.

%
\begin{table}
    \begin{center}
        \begin{tabular}{||c||c|c||c|c|c|c||c||} \hline\hline
   \ $\sqrt{s_{NN}} $ &\ $T$ &\ $\mu_B$ &\ $\overline{N_+}/\overline{N_-}$
    &\ $\omega[N_+]$ &\ $\omega[N_-]$
   &\ $\Delta[N_-,N_+]$ &\ $\langle R_{++}\rangle/\langle N_+\rangle$      \\ \hline
   \ [GeV]           & \ [MeV]  &[MeV]  &\ CE  &\ CE   &\ CE &\ CE &\ Eq.~(\ref{R-HGM1}) \\
                 \hline \hline
             \ 6.27 &\ 130.8 &\ 482.4 &\ 1.624 &\ 0.385 &\ 0.625  &\ 1.010 &\ 0.11  \\\hline
             \ 7.62 &\ 139.2 &\ 424.6 &\ 1.500 &\ 0.435 &\ 0.652  &\ 1.087 &\ 0.11  \\\hline
             \ 8.77 &\ 144.2 &\ 385.4 &\ 1.429 &\ 0.472 &\ 0.675  &\ 1.146 &\ 0.10  \\\hline
             \ 12.3 &\ 153.0 &\ 300.2 &\ 1.301 &\ 0.560 &\ 0.728  &\ 1.288 &\ 0.09  \\\hline
             \ 17.3 &\ 158.6 &\ 228.6 &\ 1.213 &\ 0.639 &\ 0.775  &\ 1.414 &\ 0.08  \\\hline
             \hline
        \end{tabular}
         \caption{The same as in Table I but in the CE. Note that relations (\ref{Rp}-\ref{Rp1}) are not fulfilled
         in the CE.
           } \label{table1-CE}
            \end{center}
\end{table}
\begin{table}
    \begin{center}
        \begin{tabular}{||c||c|c||c|c|c|c||c||} \hline\hline
   \ $\sqrt{s_{NN}} $ &\ $T$ &\ $\mu_B$ &\ $\rho[\pi^+,\pi^-]$
    &\ $\omega[\pi^+]$ &\ $\omega[\pi^-]$ &\  $\Sigma[\pi^+,\pi^-]$
    &\ $\langle R_{\pi \pi}\rangle/(\langle \pi^-\rangle + \langle \pi^+\rangle)$
    \\ \hline
   \ [GeV]           & \ [MeV]  &[MeV]  &\ CE  &\ CE &\ CE &\ CE & Eq.~(\ref{R-HGM}) \\
                 \hline \hline
             \ 6.27 &\ 130.8 &\ 482.4 &\ 0.233  &\ 0.707 &\ 0.646 &\  0.212 &\ 0.11   \\\hline
             \ 7.62 &\ 139.2 &\ 424.6 &\ 0.255  &\ 0.728 &\ 0.673  &\ 0.191 &\ 0.13   \\\hline
             \ 8.77 &\ 144.2 &\ 385.4 &\ 0.271  &\ 0.745 &\ 0.694  &\ 0.179 &\ 0.14   \\\hline
             \ 12.3 &\ 153.0 &\ 300.2 &\ 0.303  &\ 0.785 &\ 0.742  &\ 0.158 &\ 0.15   \\\hline
             \ 17.3 &\ 158.6 &\ 228.6 &\ 0.328  &\ 0.818 &\ 0.784  &\ 0.145 &\ 0.16   \\\hline
             \ 200      &\ 165.9 &\ 23.5  &\ 0.368  &\ 0.873 &\ 0.869 &\ 0.135 &\ 0.17 \\\hline
             \ 5500     &\ 166.0 &\ 0.87  &\ 0.369  &\ 0.872 &\ 0.872 &\ 0.134 &\ 0.17 \\\hline
             \hline
        \end{tabular}
         \caption{The same as in Table \ref{table2} but in the CE. Note that relations
         (\ref{R0}-\ref{R}) are not fulfilled in the CE.
                    } \label{table2-CE}
            \end{center}
\end{table}
Before presenting the UrQMD results, it is instructive
to estimate the role of the global charge conservations within the CE HGM.
These results are presented in Tables \ref{table1-CE} and \ref{table2-CE}.
All three conserved numbers, electric charge $Q$, baryonic number $B$, and strangeness $S$,
in the HGM are treated  as in the CE,
i.e. they are fixed in all microscopic states.
The hadron multiplicities are quite large ($\gg 1$) in central Pb+Pb collisions. Therefore,
the  average values of hadron multiplicities are the same in both statistical
ensembles: this means a thermodynamical equivalence of the GCE and CE.
The mean multiplicities in the GCE and CE
are approximately equal to each other when the total multiplicity
of particles carrying a conserved charge is
$N\ge 10$ (see, e.g., Ref.~\cite{CE}). This is definitely  valid for negatively and positively charged hadrons
in central Pb+Pb collisions. The mean multiplicities of produced
charged pions  are already much larger than 10
at laboratory energy higher than a few GeV per nucleon.
This is not the case for strange hadrons at  collision energies $1\div2$$A$$\,$GeV. For charmed hadrons
even at the upper SPS energy 158$A$ GeV the expected total number of charmed
hadron in central Pb+Pb collisions is of the order of unity (see Ref.~\cite{GKSG:2001}).
This, is not large enough to guarantee
a thermodynamical equivalence and equal mean multiplicities of charmed hadrons in the GCE and CE.
Note that the GCE and CE formulations considered in our paper correspond to the standard statistical mechanics.
The ideal gas system treated within the so-called Tsallis statistical mechanics
was discussed in Ref.~\cite{biro}. This approach may lead to additional physical effects which 
are not touched in the present paper. 

A thermodynamical equivalence with equal mean hadron multiplicities
in different statistical ensembles is not however extended to the particle number
fluctuations and correlations: these quantities are influenced by the global
conservation laws even in the thermodynamic limit \cite{CE}.

Note also that the CE ensemble results presented in Tables \ref{table1-CE} and \ref{table2-CE}
correspond to a case of Pb+Pb collisions, where the $Q/B$ ratio is approximately equal to 0.4.
This does not correspond to the quantum numbers in $p+p$ reactions, where $Q/B=1$.
%

%

The differences between the  HGM results in the GCE and CE
will be now illustrated by a comparison of the results presented in
Tables \ref{table2} and \ref{table2-CE}.
The values of  $\omega[\pi^+]$ and $\omega[\pi^-]$ in the CE are essentially smaller
than those in the GCE. This is the CE suppression of particle number fluctuations
due to the global charge conservation \cite{HGM}.
On the other hand, the correlation parameter $\rho[\pi^+,\pi^-]$
is larger in the CE. In the GCE, a non-zero value of $\rho[\pi^+,\pi^-]$ is due to the correlated
$\pi^+\pi^-$-pairs  coming from resonance decays. In the CE, there are
additional correlations between $\pi^+$
and $\pi^-$ numbers due to the exact charge conservation.
The differences of $\pi^+$ and $\pi^-$ number fluctuations and correlations
in CE and GCE lead
to different values of $\Sigma[\pi^+,\pi^-]$: the CE values are smaller than those in the GCE.
%
%

Note that one can use Eqs.~(\ref{R-HGM1}) and (\ref{R-HGM}) in the CE. Even more, these
equations give
the same results in the CE and GCE.
However, the relations (\ref{Rp}-\ref{Rp1}) or (\ref{R0}-\ref{R}) are
no more applicable in the CE formulation.
%
%
The matter is that  the main assumptions made in Sec.~\ref{sec-AM}
correspond to the GCE and are no more valid in the CE.

\begin{figure}[ht]
\centering
\includegraphics[width=0.49\textwidth]{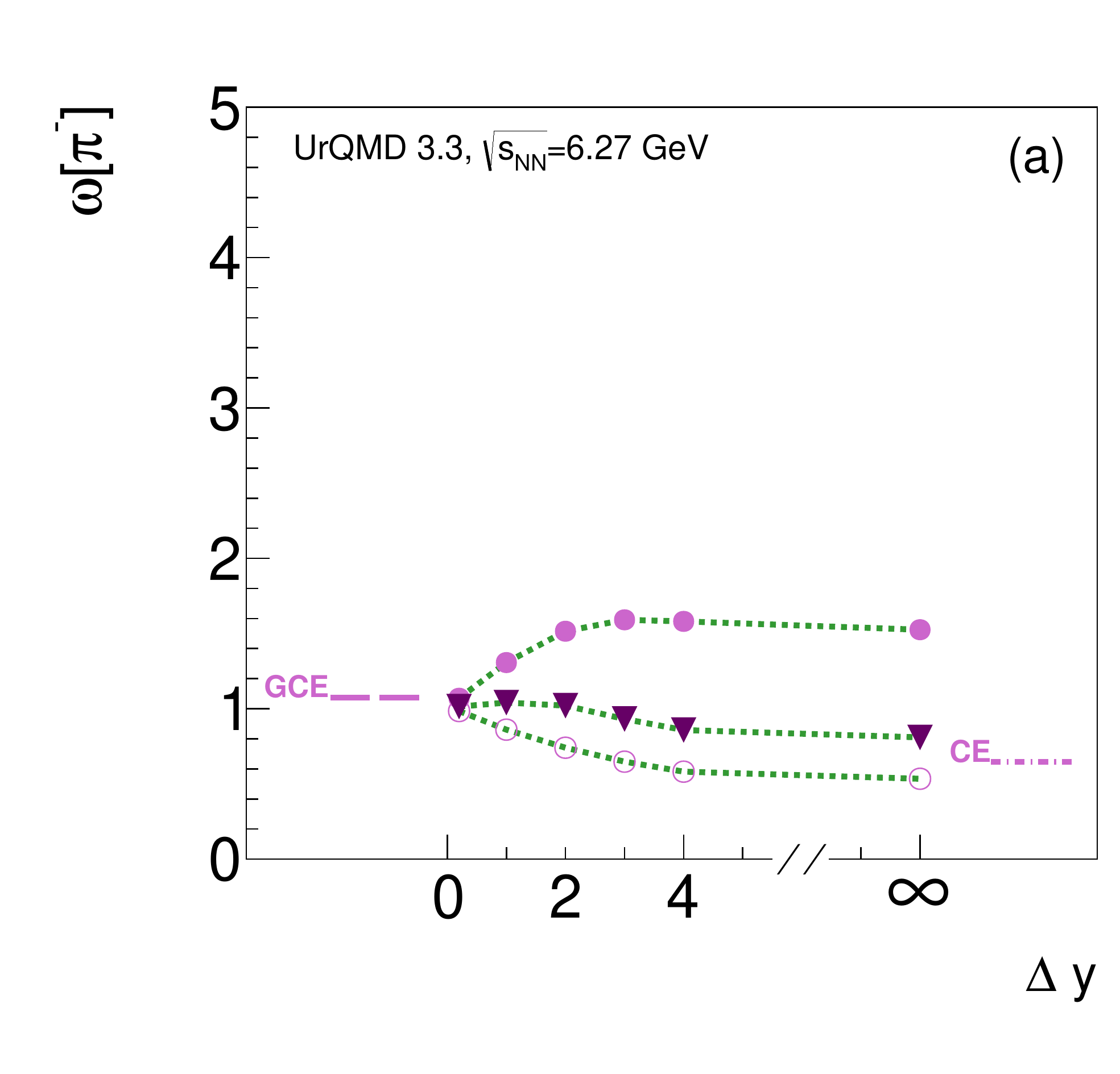}
\includegraphics[width=0.49\textwidth]{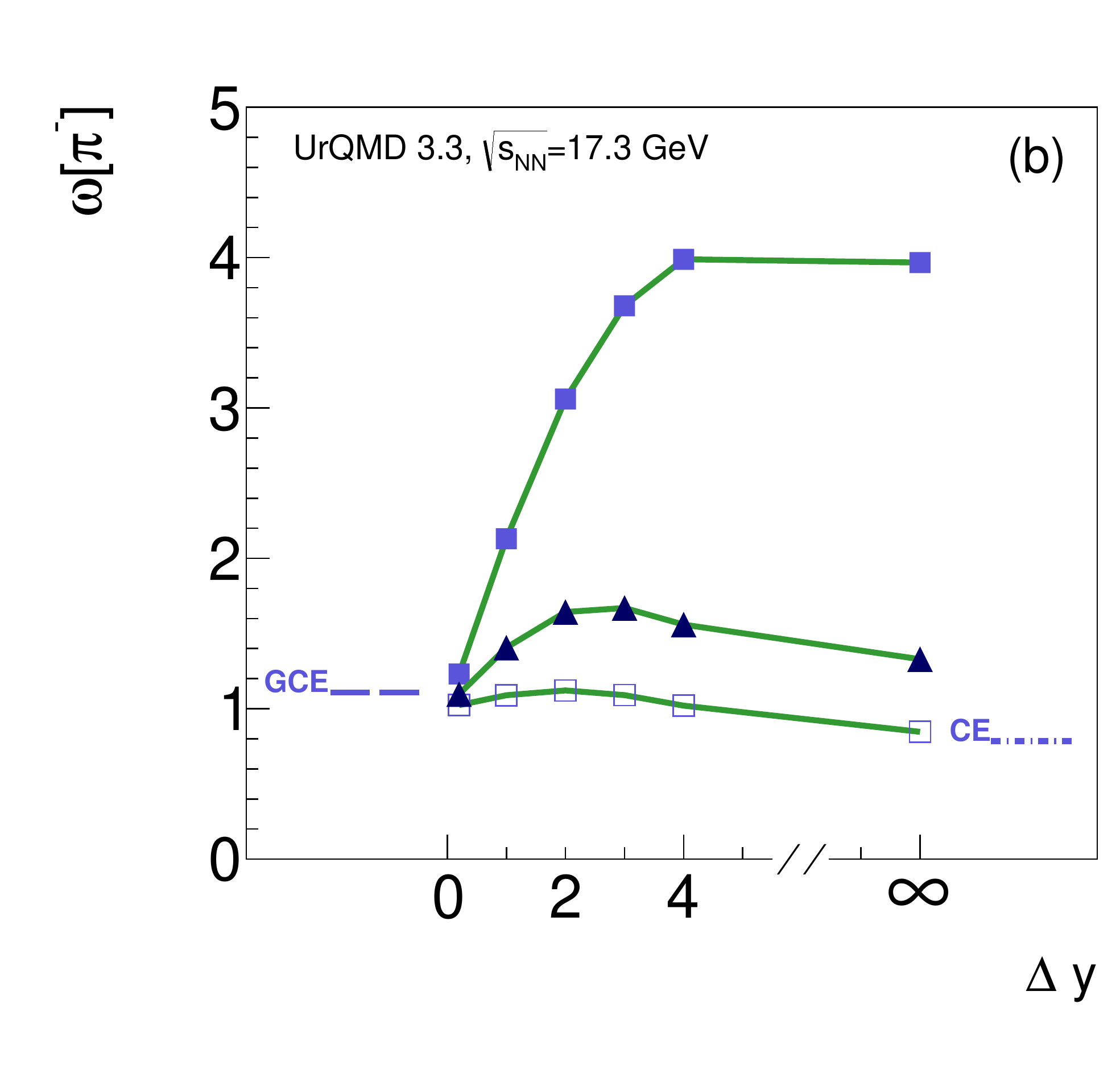}
\includegraphics[width=0.49\textwidth]{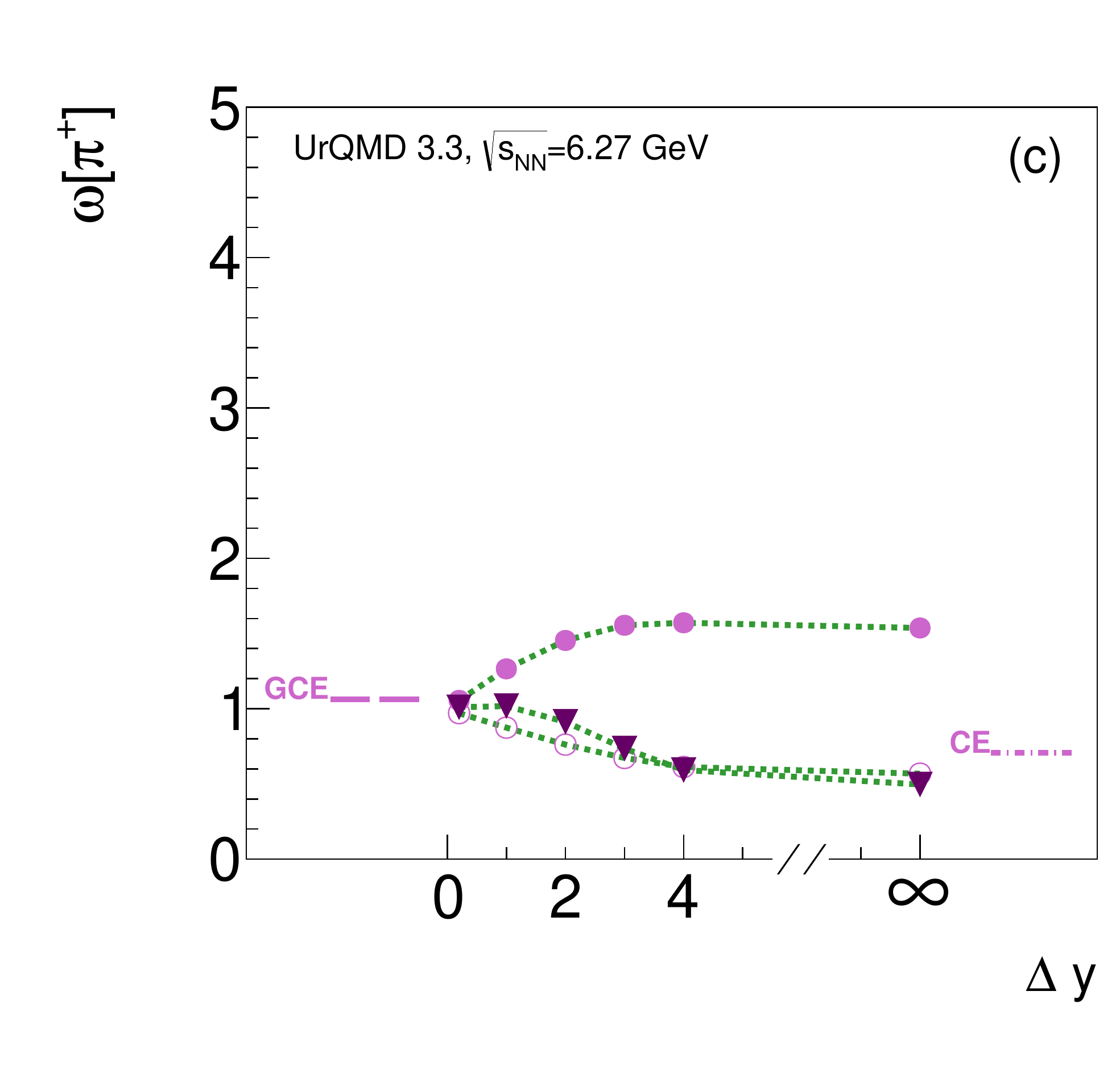}
\includegraphics[width=0.49\textwidth]{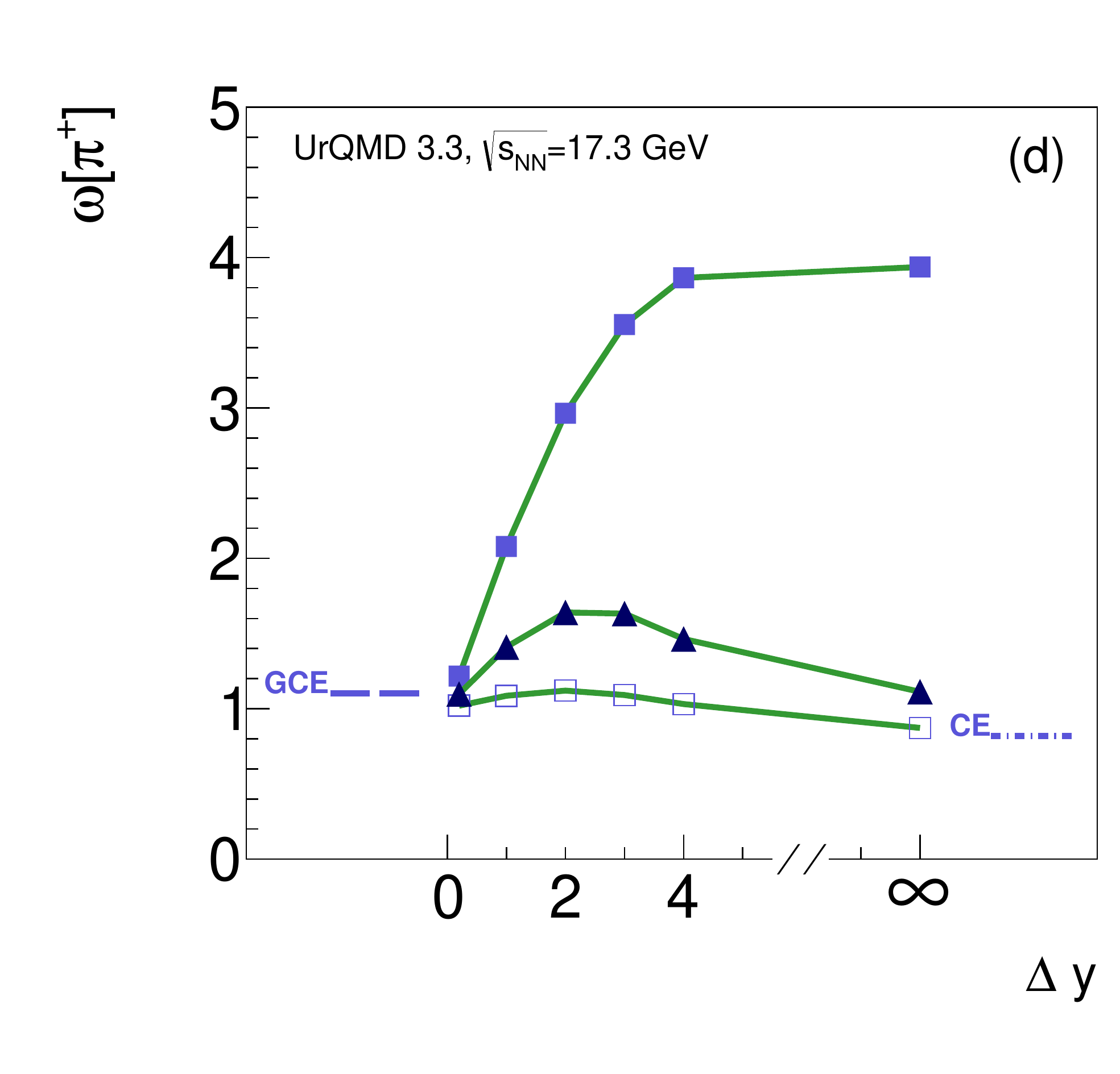}
\caption[]{The UrQMD results for the scaled variances $\omega[\pi^-]$ and $\omega[\pi^+]$
in central Pb+Pb and inelastic $p+p$ collisions for the mid-rapidity windows
$\Delta y$, i.e. the center of mass rapidities $y$ of final $\pi^+$ and $\pi^-$
satisfy the condition $-\Delta Y< y< \Delta Y$.
Full circles and squares correspond to the 5\% centrality selection and  the open ones
to the most central Pb+Pb collision events with zero impact parameter $b=0$~fm.
Triangles correspond to the UrQMD simulations of inelastic $p+p$ interactions.
The collision energies are: $\sqrt{s_{NN}}=6.27$~GeV in ({\it a}) and ({\it c}), and $\sqrt{s_{NN}}=17.3$~GeV
in ({\it b}) and ({\it d}). The horizontal
dashed and dashed-dotted lines show, respectively, the GCE (Table II) and CE (Table~IV) results
taken at the corresponding $\sqrt{s_{NN}}$.
    }
\label{fig-omega}
\end{figure}

\begin{figure}[ht]
\centering
\includegraphics[width=0.49\textwidth]{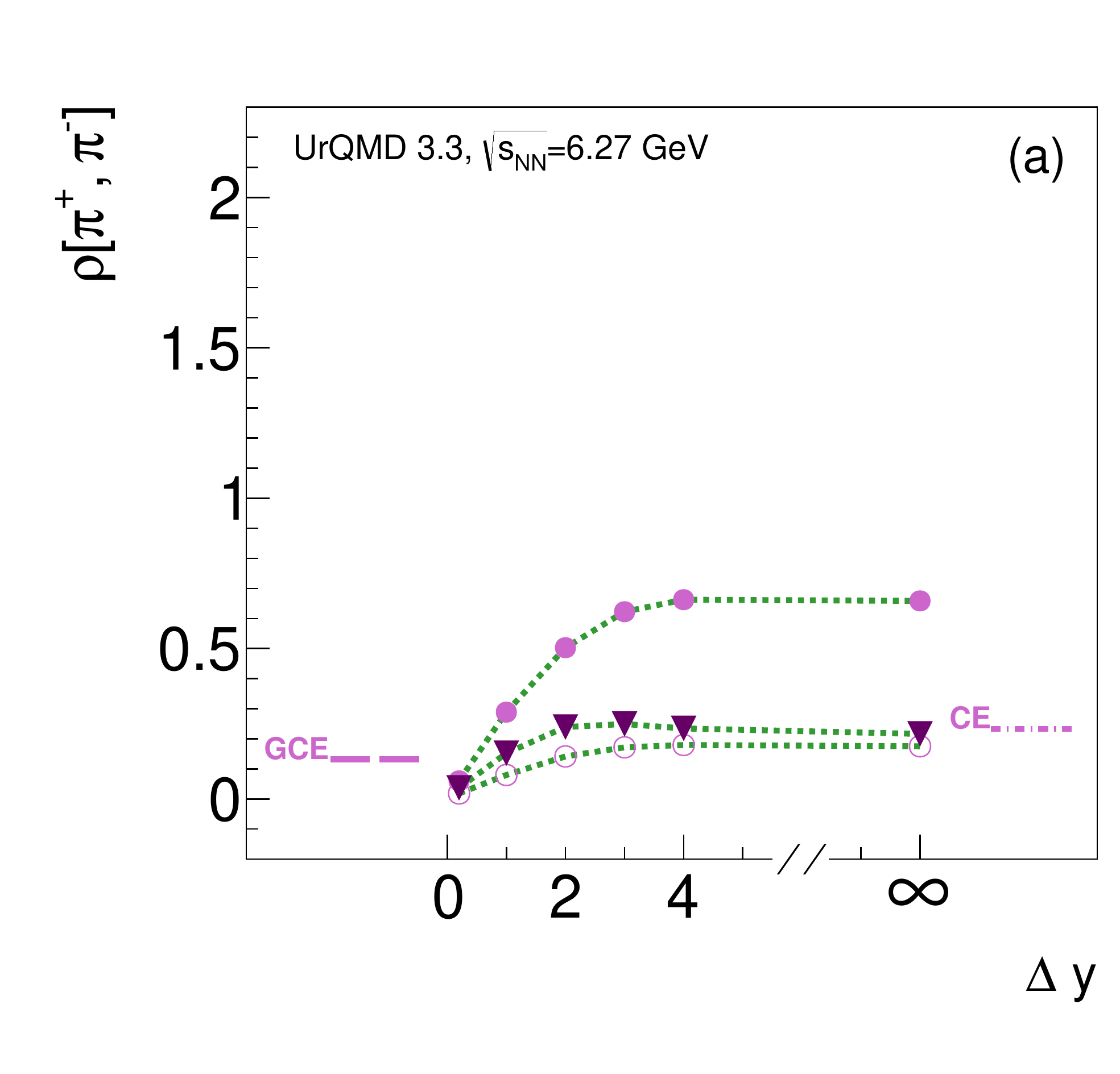}
\includegraphics[width=0.49\textwidth]{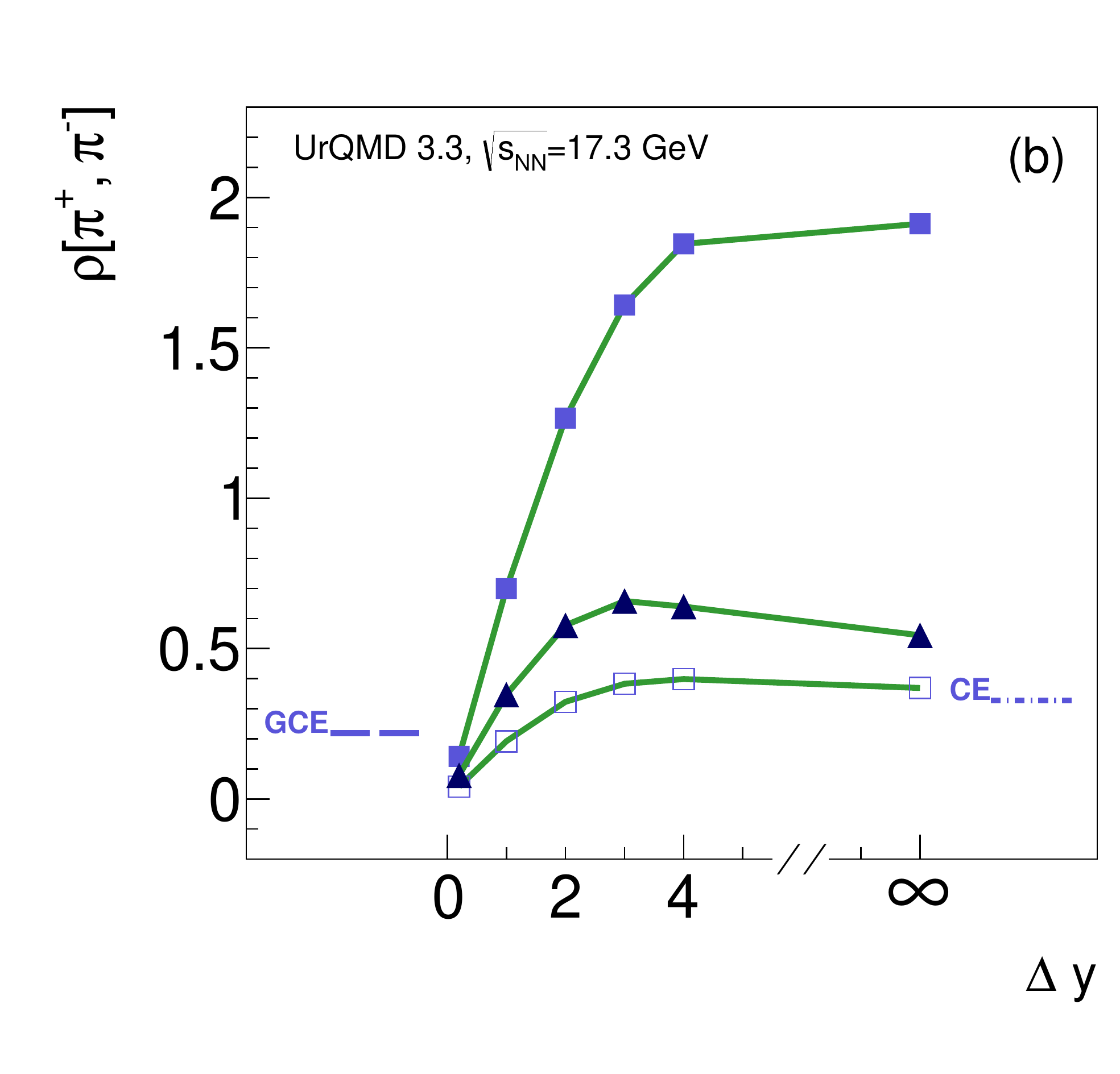}
\caption[]{The same as in Fig.~\ref{fig-omega}
but for the correlation parameter
$\rho[\pi^+,\pi^-]$.
    }
\label{fig-rho}
\end{figure}

\vspace{0.3cm}
The UrQMD values of the scaled variances $\omega[\pi^-]$, $\omega[\pi^+]$, and correlation
parameter $\rho[\pi^+,\pi^-]$ in 5\% central Pb+Pb collision
events are shown by full circles and squares in Figs.~\ref{fig-omega} and \ref{fig-rho},
respectively, as  functions of the acceptance
windows $\Delta y$. The UrQMD results for  the most central Pb+Pb collision events
with zero impact parameter, $b=0$~fm, are shown by open symbols. The triangles show
the results of the UrQMD simulations in $p+p$ reactions. The
collision energy is taken as $\sqrt{s_{NN}}=6.27$~GeV in Figs.~\ref{fig-omega} ({\it a}),({\it c}) and
\ref{fig-rho} ({\it a}), and $\sqrt{s_{NN}}=17.3$~GeV in Figs.~\ref{fig-omega} ({\it b}),({\it d}) and
\ref{fig-rho} ({\it b}).
The windows at the center of mass mid-rapidity are taken as
$\Delta y= 0.2$, 1, 2, 3, 4, and $\infty$.
A symbol $\infty$ denotes the case when all final state particles are detected
(i.e. a full $4\pi$-acceptance).

Note that the UrQMD model does not assume
any, even local, thermal and/or chemical equilibration.
Therefore, a connection between the  UrQMD and HGM results for particle number fluctuations
and correlations is a priori unknown.

For the 5\% most central Pb+Pb collisions the scaled variance $\omega[\pi^-]$, shown in Fig.~\ref{fig-omega} ({\it a}) and ({\it b}), increases with $\Delta y$.  As seen from Fig.~\ref{fig-omega} ({\it b}),
this increase is rather strong at high collision energy:
at large $\Delta y$,
the value of  $\omega[\pi^-]$ becomes
much larger than the HGM results in both the
CE and GCE. The behavior of $\omega[\pi^+]$, shown in Fig.~\ref{fig-omega} ({\it c}) and ({\it d}),
is rather similar
to that of $\omega[\pi^-]$.

The correlation
parameter $\rho[\pi^+,\pi^-]$ is presented in Fig.~\ref{fig-rho} ({\it a}) and ({\it b}).
Selecting within the UrQMD simulations the most central Pb+Pb collision
events with zero impact parameter $b=0$~fm, one finds essentially smaller
values of $\omega[\pi^-]$, $\omega[\pi^+]$ (open symbols in Fig.~\ref{fig-omega}), and $\rho[\pi^+,\pi^-]$.
This means that in the 5\% centrality bin of Pb+Pb collision events
large fluctuations of the number of
nucleon participants (i.e., the volume fluctuations) are present.
These volume fluctuations  produce large additional contributions
to the scaled variances of pions and to the correlation parameter $\rho[\pi^+,\pi^-]$.
These contributions were presented as the separate terms
in Eqs.~(\ref{om1V}-\ref{12V}). They become more and more important with increasing
collision energy. This is due to an increase  of the pion number
density (or, similarly, the number of pions per participating nucleon) with increasing collision energy.
However, one hopes that these volume fluctuations will be canceled out to a large
extent when they are combined in the strongly intensive measures.
Note also that the UrQMD results for $\omega[\pi^-]$, $\omega[\pi^+]$, and $\rho[\pi^+,\pi^-]$
in inelastic $p+p$ collisions,
shown in Figs.~\ref{fig-omega} and \ref{fig-rho} by triangles, are qualitatively similar to those
in Pb+Pb collisions at $b=0$~fm.

\vspace{0.3cm}

\begin{figure}[ht]
\centering
\includegraphics[width=0.495\textwidth]{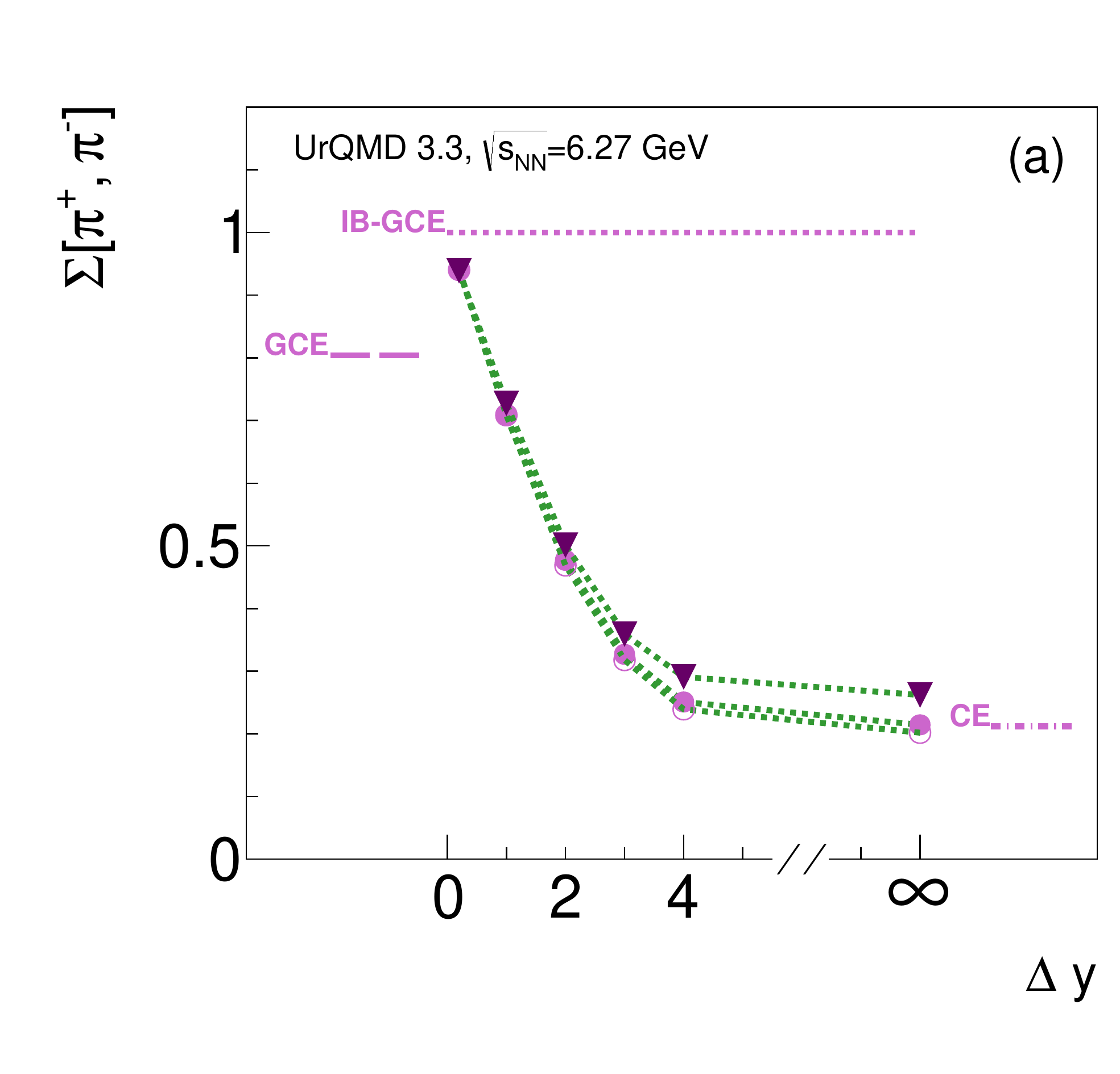}
\includegraphics[width=0.495\textwidth]{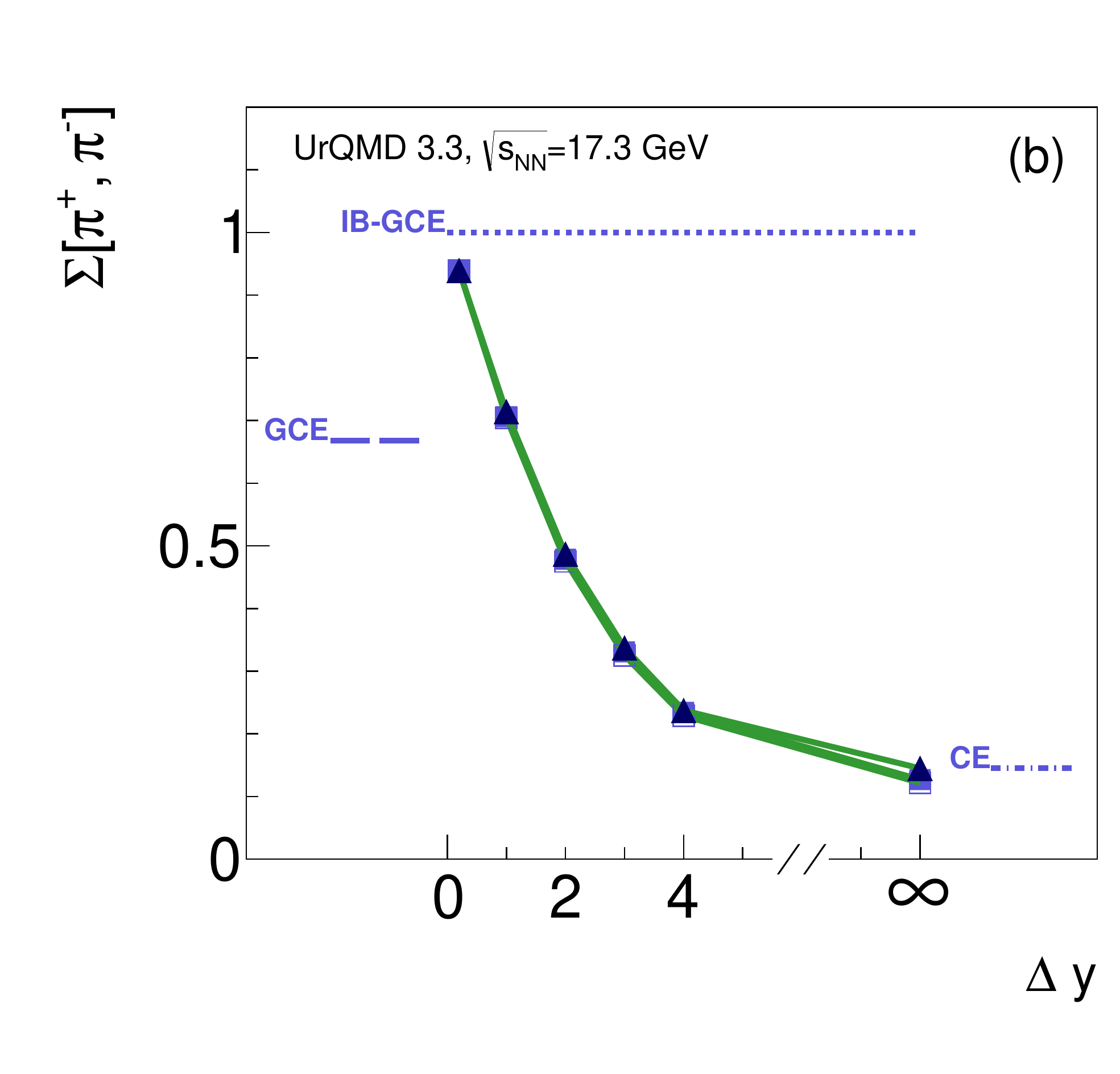}
\caption[]{The same as in Figs.~\ref{fig-omega} and \ref{fig-rho} but
for $\Sigma[\pi^+,\pi^-]$. The horizontal dotted line shows the IB-GCE result
$\Sigma[\pi^+,\pi^-]=1$.
 }
\label{fig-sigma}
\end{figure}

The UrQMD results for $\Sigma[\pi^+,\pi^-]$ in Pb+Pb collisions
at $\sqrt{s_{NN}}=6.27$~GeV and 17.3~GeV are presented in Fig.~\ref{fig-sigma} ({\it a}) and ({\it b}),
respectively, as a function of the acceptance window
$\Delta y$ at mid-rapidity.

%
%
%
In contrast to the results shown in Figs.~\ref{fig-omega} and
\ref{fig-rho}, both centrality selections in Pb+Pb collisions (5\% centrality bin and $b=0$~fm)
lead to very similar results for $\Sigma[\pi^+,\pi^-]$ shown in Fig.~\ref{fig-sigma}.
This means that the measure $\Sigma[\pi^+,\pi^-]$ has the strongly intensive properties,
at least in the UrQMD
simulations. The UrQMD results in p+p
reactions are close to those in  Pb+Pb ones.

The GCE and CE results from Tables \ref{table2} and \ref{table2-CE}
are presented in Figs.~\ref{fig-omega}-\ref{fig-sigma}
by the horizontal dashed and dashed-dotted lines, respectively.
The UrQMD results for $\Sigma[\pi^+,\pi^-]$, presented in Fig.~\ref{fig-sigma},
demonstrate a strong dependence on
the size of rapidity window $\Delta y$.
At  $\Delta y=1$ these results
are close to those of the GCE HGM.
On the other hand,
with increasing $\Delta y$ the role of exact charge conservation becomes more and
more important.
%
From Fig.~\ref{fig-sigma},  one observes
that the UrQMD values  of $\Sigma[\pi^+,\pi^-]$ at large $\Delta y$
are close to the CE results.

As seen from Figs.~\ref{fig-omega} and \ref{fig-rho},
a similar correspondence between the UrQMD results for $\omega[\pi^-]$,
$\omega[\pi^+]$, and $\rho[\pi^+,\pi^-]$ in Pb+Pb collisions at $b=0$~fm
and their GCE and CE values  is approximately valid. However,
this is not the case for  the 5\% most central Pb+Pb events. In that centrality
bin the volume fluctuations give the dominant contributions to
$\omega[\pi^-]$,
$\omega[\pi^+]$, and $\rho[\pi^+,\pi^-]$  for large $\Delta y$.

For very small acceptance, $\Delta y \ll 1$,
one expects an approximate validity of the Poisson distribution for
any type of the detected particles. Their scaled variances are then
close to unity, i.e., $\omega[\pi^-]\cong \omega[\pi^+] \cong 1$.
Particle number correlations,
due to both the resonance decays and the global charge conservation,
become negligible, i.e., $\rho[\pi^+,\pi^-]\ll 1$. Therefore, at very small
$\Delta y$ the IB-GCE results should be valid.
These expectations are indeed supported by the UrQMD results at $\Delta y=0.2$
presented in Figs.~\ref{fig-omega} and \ref{fig-rho}.
Therefore, one expects $\Sigma[\pi^+,\pi^-]\rightarrow 1$ at $\Delta y\rightarrow 0$.
This expectation is also valid, as seen from the UrQMD results at $\Delta y=0.2$
presented in Fig.~\ref{fig-sigma}.

\section{Summary}\label{sum}

In this paper we use the {\it strongly intensive} measures
$\Delta$ and $\Sigma$
and analyze the effects of resonance decays
for the particle number fluctuations and correlations.
Two examples for which the event by event fluctuations of hadron multiplicities
are rather sensitive to the abundances of
resonances at the chemical freeze-out are discussed:
resonance decays to two positively charged particles, like $\Delta^{++}\rightarrow p+ \pi^+$,
and to $\pi^+\pi^-$-pair, like $\rho^0\rightarrow \pi^-+\pi^+$.
Simple analytical formulation demonstrates that
the resonance abundances, which are difficult to be measured by other methods,
can be found by measuring the fluctuations and correlations of the numbers of stable hadrons.
The  grand canonical ensemble calculations within the hadron-resonance gas model
support these physical results.

The ultra-relativistic quantum molecular dynamics
model is used in Pb+Pb and  $p+p$ collisions at the SPS energies
to illustrate the role of centrality selection, limited acceptance,
and global charge conservation.  A crucial importance of the size
of the rapidity window for the accepted particles is emphasized.
It should be larger than unity
for a simultaneous hit into this rapidity window
of both  correlated hadrons,  e.g., $\pi^+$ and $\pi^-$, from resonance decays.
However, if this window of accepted particles is comparable to the whole rapidity interval
the restrictions of the exact global charge conservations become important. This is illustrated
by the canonical ensemble calculations when the conserved charges
are fixed for all microscopic states. The global charge conservation influences
the particle number fluctuations  and introduces additional
correlations between numbers of different particle species. Thus, a connection
between the fluctuation measures and the resonance abundances becomes more complicated.
The high energy RHIC and LHC accelerators look therefore preferable for these investigations:
one can use a large enough rapidity interval (comparing to unity) which will be only a small
part of the whole system.

\begin{acknowledgments}
We thank Michael Hauer for providing us the THERMUS code extended for fluctuations.
We are indebted to the authors of the UrQMD model for the use of their code in our analysis.
We are thankful to Marek Ga\'zdzicki  for
fruitful discussions and comments.
V.B. was supported in part by the Polish National Science Center grant
with decision No. DEC-2012/06/A/ST2/00390. The work of M.I.G.  was
supported by the National Academy of Sciences of Ukraine,
research Grant ZO-2-1/2014, and
by the State Agency of Science, Innovations and
Informatization of Ukraine contract F58/384-2013.
The work of K.G. was supported by the National Science Center,
Poland grant DEC-2011/03/B/ST2/02617 and grant 2012/04/M/ST2/00816.
\end{acknowledgments}



\end{document}